\documentclass[aps,
reprint,
groupedaddress,
]{revtex4-2}
\usepackage{hyperref}
\usepackage[pdftex]{graphicx}
\usepackage{bm}
\usepackage{amsmath,amssymb,amsfonts,mathtools}
\usepackage{physics}
\usepackage{xcolor}

\begin{document}

\title{Revisiting the surface density of states of midgap Andreev edge states}

\author{Gota Sato}
\author{Yasushi Nagato}
\author{Seiji Higashitani}
\email[]{ultp@hiroshima-u.ac.jp}
\affiliation{
  Graduate School of Advanced Science and Engineering,
  Hiroshima University,
  1-7-1 Kagamiyama, Higashi-Hiroshima City, Hiroshima, 739-8521, Japan
}

\date{\today}

\begin{abstract}
We revisit the effect of surface roughness on midgap Andreev edge states (MAES) in $p$- and $d$-wave superconductors. For a perfectly specular surface, MAES form a flat band at the Fermi energy, which manifests as a sharp midgap peak in the surface density of states (SDOS). Previous theoretical studies have shown that MAES in $p$- and $d$-wave superconductors respond markedly differently to surface roughness. In the $d$-wave state, diffuse surface scattering significantly broadens the midgap peak in the SDOS, accompanied by the emergence of a V-shaped structure centered at the Fermi energy. In contrast, the midgap peak in the $p$-wave state remains robust against diffuse scattering. In this work, we clarify the physical origin of this contrasting behavior. A key aspect of our analysis is that the flat band in the $d$-wave state consists of two distinct types of MAES modes. We show that inter-mode diffuse scattering leads to substantial broadening of the midgap peak and to the formation of the V-shaped structure. By contrast, the robustness of MAES in the $p$-wave state arises from the presence of a single MAES mode in the flat band. These results provide new insight into the response of MAES to surface roughness.
    
\end{abstract}

\maketitle

\section{Introduction \label{sec:Introduction}}

Midgap Andreev edge states (MAES) have attracted considerable attention since their existence was first predicted in the context of $d$-wave superconductivity in high-$T_c$ cuprates \cite{Hu_PRL1994}. MAES arise as a direct consequence of anisotropic Cooper pairing, as realized in $p$- and $d$-wave superconducting (SC) states. Specifically, they form at surfaces where the SC gap function changes sign along a quasiparticle reflection trajectory \cite{Hu_PRL1994,Ohashi_JPSJ1996}. Because MAES constitute a flat band at the Fermi energy, they appear in the surface density of states (SDOS) as a pronounced zero-energy peak with large spectral weight. This feature distinguishes MAES from other types of edge states, such as chiral and helical edge states, whose energy bands disperse within the SC gap \cite{Tanaka_JPSJ2012,Mizushima_JPSJ2016}. 

The flat MAES band leads to observable phenomena, including a zero-bias peak in tunneling conductance spectra \cite{Tanaka_PRL1995,Kashiwaya_RPP2000} and a low-temperature anomaly in the critical Josephson current \cite{Barash_PRL1996}. Furthermore, it has been shown that the large zero-energy SDOS can induce Fermi surface instability, resulting in a surface state that spontaneously breaks time-reversal symmetry \cite{Matsumoto_I_JPSJ1995,*Matsumoto_II_JPSJ1995,*Matsumoto_III_JPSJ1995,Higashitani_JPSJ1997,Fogelstrom_PRL1997,Sigrist_PTP1998,Lofwander_PRB2000,Vorontsov_PRL2009,Hakansson_NatPhys2015,Holmvall_NatCommun2018,Miyawaki_PRB2018}. This superconducting state, characterized by locally broken time-reversal symmetry along surfaces hosting MAES, has been proposed as a novel phase in $d$-wave superconductors.

In general, surface properties of superconductors and superfluids are highly sensitive to boundary conditions \cite{Matsumoto_JPSJ1995,Yamada_JPSJ1996,Barash_PRL1996,Luck_PRB2001,Bakurskiy_PRB2014,Miyawaki_PRB2018,Nagato_JLTP1998,Vorontsov_PRB2003,Nagai_JPSJ2008,Nagato_JPSJ2011}. However, most theoretical studies of MAES to date have relied on idealized surface models that assume purely specular reflection. A pioneering investigation of diffuse surface scattering effects on MAES was conducted by Matsumoto and Shiba \cite{Matsumoto_JPSJ1995}. Using the randomly rippled wall model \cite{Chaplik_JETP1969,Falkovskii_JETP1970}, they calculated the SDOS for a $d$-wave SC state in cuprates and demonstrated that diffuse scattering at a rough surface significantly broadens the midgap SDOS peak. Similar conclusions were reached by Yamada et al.\,\cite{Yamada_JPSJ1996} and L\"uck et al.\,\cite{Luck_PRB2001}, who employed an alternative approach in which boundary conditions are parameterized by an $S$-matrix.

The pronounced broadening of the SDOS peak indicates a suppression of anomalous surface and interface phenomena associated with MAES \cite{Barash_PRL1996,Miyawaki_PRB2018}. In contrast, MAES in $p$-wave superconductors have been shown to be remarkably robust against surface roughness \cite{Nagato_JLTP1998,Bakurskiy_PRB2014}. This robustness has been qualitatively attributed to the presence of odd-frequency $s$-wave Cooper pairs associated with MAES in the $p$-wave state \cite{Higashitani_PRB2012,Bakurskiy_PRB2014,Miyawaki_PRB2018}.

In this paper, we revisit the problem of surface roughness in MAES. We analyze the SDOS in $p$- and $d$-wave SC states using a quasiclassical framework that we recently developed to investigate SDOS broadening in two-dimensional (2D) chiral superconductors \cite{Higashitani_PRB2024}. Here, we reformulate and extend this theory to nonchiral $p$- and $d$-wave superconductors. Our objectives are twofold. First, we aim to clarify why MAES exhibit markedly different responses to surface roughness in $p$- and $d$-wave superconductors. Second, we address the origin of the non-Lorentzian SDOS observed in the $d$-wave SC state. As reported by Matsumoto and Shiba \cite{Matsumoto_JPSJ1995}, the broadening of the midgap peak is accompanied by the emergence of a V-shaped structure centered at zero energy. The physical origin of this characteristic feature has remained unclear, and we seek to resolve this issue.

The remainder of this paper is organized as follows. Section \ref{sec:model-method} introduces the theoretical model and outlines the quasiclassical framework for MAES. Section \ref{sec:SDOS-specular} examines the specular surface case and establishes the basis for analyzing diffuse scattering effects. Section \ref{sec:SDOS-rough} discusses the impact of diffuse scattering on the SDOS of MAES in $p$- and $d$-wave superconductors. The final section presents our conclusions.

\section{Model and method}
\label{sec:model-method}

We consider a two-dimensional (2D) superconductor occupying the region $y > 0$, with a surface located along the $x$ axis. The surface may exhibit atomic-scale irregularities, although it is macroscopically flat. For simplicity, the Fermi surface is assumed to be isotropic (circular). We focus on superconducting (SC) states characterized by the gap function
\begin{equation}
  \Delta_\alpha(y,\phi) = \alpha \Delta(y) \sin(m\phi),
  \label{eq:gap-def}
\end{equation}
where $\Delta(y)$ is a real amplitude and $m$ is an integer. We assume $\Delta(\infty) > 0$. The cases $m = 1$ and $m = 2$ correspond to $p$-wave and $d$-wave SC states, respectively. The anisotropy of the gap function in $k$-space is described by the angle $\phi = \arccos(k_x/k)$ ($0 < \phi < \pi$) and the directional index $\alpha = {\rm sgn}(k_y)$. Because the gap function in Eq.\ \eqref{eq:gap-def} satisfies the symmetry relation $\Delta_+(y,\phi) = -\Delta_-(y,\phi)$, MAES are formed for all angles $\phi$ \cite{Hu_PRL1994,Ohashi_JPSJ1996}.

We apply the quasiclassical theory of superconductivity to this model system. A central quantity in this framework is the quasiclassical Green's function $\hat{g}_\alpha$, which satisfies the Eilenberger equation \cite{Eilenberger_ZPhys1968,Serene_PR1983}
\begin{equation}
  \alpha i \hbar v_y \partial_y \hat{g}_\alpha(y)
  = [\hat{g}_\alpha(y),\ \hat{h}_\alpha(y,\phi,\varepsilon)],
  \label{eq:Eilen}
\end{equation}
together with the normalization condition
\begin{equation}
  \hat{g}_\alpha^2(y) = -1.
  \label{eq:nor-con}
\end{equation}
Here, $v_y$ is the component of the Fermi velocity normal to the surface, and $\varepsilon$ is a complex energy variable. The matrix $\hat{h}_\alpha$ is defined as
\begin{equation}
  \hat{h}_\alpha(y,\phi,\varepsilon) =
  \begin{pmatrix}
    \varepsilon & \Delta_\alpha(y,\phi) \\
    -\Delta_\alpha^*(y,\phi) & -\varepsilon \\
  \end{pmatrix}.
\end{equation}
The diagonal elements of $\hat{g}_\alpha(y)$ contain information about the quasiparticle density of states. In this study, we focus on the surface density of states (SDOS), normalized to unity in the normal state, given by
\begin{equation}
  n(\phi, E) = {\rm Im}
  \left[
    \frac{1}{2}{\rm Tr}\sum_{\alpha=\pm}
    \frac{\hat{\rho}_3 \hat{g}^R_\alpha(0)}{2}
  \right],
  \label{eq:SDOS-def}
\end{equation}
where $\hat{\rho}_3$ is the third Pauli matrix, and the superscript $R$ denotes the retarded Green's function evaluated at $\varepsilon = E + i0^+$. The off-diagonal elements yield the self-consistency equation for the gap function $\Delta_\alpha(y,\phi)$ (see Refs.\ \cite{Zhang_PLA1985,Kieselmann_PRB1987,Yamada_JPSJ1996,Nagato_JLTP1998} for details).

The surface boundary condition for $\hat{g}_\alpha(y)$ can be derived by introducing an $S$-matrix for electrons at the Fermi level in the normal state and subsequently transforming from the wave-function formalism to the Green's-function formalism \cite{Nagato_JLTP1996}. The simplest case corresponds to specular reflection, for which $\mathcal{S}_{K,K'} = -\delta_{K,K'}$, where $K$ denotes the wavevector component parallel to the surface (i.e., $k_x$ in the present 2D system). This yields the well-known specular boundary condition
\begin{equation}
  \hat{g}_+(0) = \hat{g}_-(0) \equiv \hat{g}_{S}.
\end{equation}

\begin{figure}
  \centering
  \includegraphics[scale=0.85]{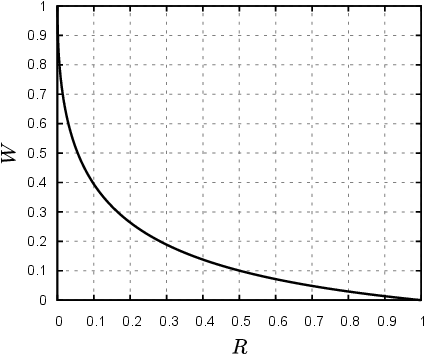}
  \caption{Roughness parameter $W$ as a function of the specular reflection probability $R$.}
    \label{fig:W-R}
\end{figure}

To incorporate diffuse surface scattering, we adopt the random $S$-matrix model proposed by Nagato et al.~\cite{Nagato_JLTP1996},
\begin{equation}
  \mathcal{S}_{K,K'}
  = -\left(\frac{1 - i\eta}{1 + i\eta}\right)_{K,K'}.
\end{equation}
This parameterization ensures unitarity of the $S$-matrix for any Hermitian matrix $\eta$. Treating $\eta$ as a random variable with Gaussian statistics, Nagato et al.\ derived the boundary condition
\begin{equation}
  \hat{g}_+(0) =
  \frac{1 + i\hat{\gamma}}{1 - i\hat{\gamma}}\,
  \hat{g}_-(0)
  \frac{1 - i\hat{\gamma}}{1 + i\hat{\gamma}},
  \label{eq:bcon-rough}
\end{equation}
which incorporates diffuse scattering effects through $\hat{\gamma}$:
\begin{gather}
  \hat{\gamma} = 2W
  \left<
    \frac{1}{\hat{g}_S^{\,-1} - \hat{\gamma}}
  \right>_\phi,
  \label{eq:gamma-W}
  \\
  \left<\cdots\right>_\phi
  = \frac{\sum_{K}(\cdots)}{\sum_{K}1}
  = \frac{1}{2}\int_{0}^{\pi} d\phi\, \sin\phi\, (\cdots).
  \label{eq:phi-ave}
\end{gather}
In Eq.\ \eqref{eq:gamma-W}, $W$ is a roughness parameter ranging from 0 to 1 \cite{Yamada_JPSJ1996,Nagato_JLTP1996,Miyawaki_PRB2018,Higashitani_PRB2024}. The limits $W = 0$ and $W = 1$ correspond to purely specular and fully diffuse scattering, respectively, where the latter implies isotropic scattering of incident electrons. The parameter $W$ is related to the specular reflection probability $R$ as $W = (1 - \sqrt{R})/(1 + \sqrt{R})^2$ \cite{Miyawaki_PRB2018,Higashitani_PRB2024} (Fig.\ \ref{fig:W-R}).

Recently, we employed the above random $S$-matrix framework to investigate the influence of diffuse scattering on edge states in 2D chiral superconductors \cite{Higashitani_PRB2024}. In the following, we outline the corresponding theoretical framework adapted to the nonchiral superconductors considered in this study.

The quasiclassical Green's function can be constructed using the Andreev amplitudes $\varphi_{\alpha}^{l}(y)$, which satisfy
\begin{equation}
  \big(-\alpha i \hbar v_y \partial_y \hat{\rho}_{3}
    + \alpha \Delta(y) \sin(m\phi) \hat{\rho}_{1}\big)
  \varphi_\alpha^{l}(y) = \varepsilon \varphi_\alpha^{l}(y),
  \label{eq:Andreev}
\end{equation}
where the superscript $l = g,d$ denotes the growing ($\varphi_{\alpha}^{g}$) and decaying ($\varphi_{\alpha}^{d}$) solutions as $y \to \infty$. Their asymptotic forms are
\begin{gather}
  \varphi_{\alpha}^{g}(y \to \infty) \propto
  \begin{pmatrix}
    \alpha \Delta_{\phi} \\
    \varepsilon + \alpha i \Omega \\
  \end{pmatrix}
  e^{+\Omega y/\hbar v_{y}}, \\
  \varphi_{\alpha}^{d}(y \to \infty) \propto
  \begin{pmatrix}
    \varepsilon + \alpha i \Omega \\
    \alpha \Delta_{\phi} \\
  \end{pmatrix}
  e^{-\Omega y/\hbar v_{y}},
\end{gather}
where
\begin{equation}
  \Omega = \sqrt{\Delta_{\phi}^{2} - \varepsilon^{2}}, \quad
  \Delta_{\phi} = \Delta(\infty) \sin(m\phi).
\end{equation}

In this model, the Andreev amplitudes satisfy the symmetry relation $\varphi_{+}^{l}(y) \propto \hat{\rho}_{2} \varphi_{-}^{l}(y)$. Taking this particle-hole symmetry into account, we introduce the parameterizations
\begin{gather}
  \varphi_+^d(y) = u_{+}^{d}(y)
  \begin{pmatrix} 1 \\ D(y) \end{pmatrix}
  \propto \hat{\rho}_{2} \varphi_-^d(y), \\
  \varphi_-^g(y) = u_{-}^{g}(y)
  \begin{pmatrix} 1 \\ F(y) \end{pmatrix}
  \propto \hat{\rho}_{2} \varphi_+^g(y).
\end{gather}
The functions $D(y)$ and $F(y)$ satisfy the Riccati-type differential equations
\begin{gather}
  i\hbar v_{y} \partial_{y} D
  = 2\varepsilon D - \Delta(y)\sin(m\phi) (1 + D^{2}),
  \label{eq:R-D} \\
  i\hbar v_{y} \partial_{y} F
  = -2\varepsilon F - \Delta(y)\sin(m\phi) (1 + F^{2}).
  \label{eq:R-F}
\end{gather}

The quasiclassical Green's functions $\hat{g}_{\alpha}(y)$ satisfying Eqs.\ \eqref{eq:Eilen} and \eqref{eq:nor-con} can be expressed in terms of $D(y)$ and $F(y)$ as
\begin{align}
  \hat{g}_+(y)
  &= \frac{2i\hat{\rho}_{3}}{1 + D(y)F(y)}
  \begin{pmatrix} 1 \\ D(y) \end{pmatrix}
  \begin{pmatrix} 1 & -F(y) \end{pmatrix} - i,
  \label{eq:gp-DF} \\
  \hat{g}_-(y)
  &= \hat{g}_+^T(y).
  \label{eq:gm-DF}
\end{align}
Thus, the boundary-value problem for $\hat{g}_{\alpha}(y)$ reduces to solving for $D(y)$ and $F(y)$ with the boundary conditions
\begin{gather}
  D(\infty)
  = \frac{\Delta_{\phi}}{\varepsilon + i\Omega},
  \label{eq:bcon-D} \\
  \begin{pmatrix} 1 \\ F(0) \end{pmatrix}
  \propto
  \hat{\rho}_{3}
  \frac{1 - i\hat{\gamma}}{1 + i\hat{\gamma}}
  \hat{\rho}_{3}
  \begin{pmatrix} 1 \\ D(0) \end{pmatrix}.
  \label{eq:bcon-F}
\end{gather}
Equations \eqref{eq:R-D}--\eqref{eq:bcon-F} form a closed set for determining the quasiclassical Green's function in this system.

In the specular limit, the condition $F(0) = D(0)$ holds. From Eq.\ \eqref{eq:gp-DF}, the surface Green's function is then given by
\begin{equation}
  \hat{g}_{S}
  = \hat{\rho}_{3}\left(G_{1}\ketbra{1}{1} + G_{2}\ketbra{2}{2}\right),
  \label{eq:gs-braket}
\end{equation}
where
\begin{equation}
  G_{1} = i\,\frac{1 + iD(0)}{1 - iD(0)}, \quad
  G_{2} = i\,\frac{1 - iD(0)}{1 + iD(0)},
\end{equation}
and
\begin{equation}
  \ket{1} = \frac{1}{\sqrt{2}}
  \begin{pmatrix} 1 \\ -i \end{pmatrix}, \quad
  \ket{2} = \frac{1}{\sqrt{2}}
  \begin{pmatrix} 1 \\ i \end{pmatrix}
\end{equation}
are eigenvectors of $\hat{\rho}_{2}$. Notably, the Andreev equation at $\varepsilon = 0$ admits eigenstates proportional to $\ket{1}$ and $\ket{2}$, as can be verified from Eq.\ \eqref{eq:Andreev} after multiplication by $\hat{\rho}_{3}$ from the left. This indicates that $G_{1}$ and $G_{2}$ encode information about two distinct MAES modes proportional to $\ket{1}$ and $\ket{2}$, respectively (see also Sec.\ \ref{sec:SDOS-specular}).

The matrix $\hat{\gamma}$ has a structure analogous to $\hat{g}_S$:
\begin{equation}
  \hat{\gamma} = \hat{\rho}_3
  \left(S_1\ketbra{1}{1} + S_2\ketbra{2}{2}\right).
  \label{eq:gamma-S}
\end{equation}
Substituting Eq.\ \eqref{eq:gamma-S} into Eq.\ \eqref{eq:gamma-W} yields
\begin{equation}
  S_1 = 2W \ev{G_1^W}_{\phi}, \quad
  S_2 = 2W \ev{G_2^W}_{\phi},
  \label{eq:S12-G}
\end{equation}
where
\begin{equation}
  G_1^W = \frac{G_1}{1 - G_1 S_2}, \quad
  G_2^W = \frac{G_2}{1 - G_2 S_1}.
  \label{eq:G12W-def}
\end{equation}
From these expressions, the SDOS can be written in the form (see Appendix)
\begin{subequations}
  \label{eq:SDOS-formula}
\begin{gather}
  n(\phi,E) = n_1(\phi,E) + n_2(\phi,E),
  \label{eq:SDOS-cG-mat} \\
  n_j(\phi,E) = \frac{1}{2}{\rm Im}(\mathcal{G}_j^R)
  \quad (j = 1,2), \\
  \mathcal{G}_1 = \frac{G_1 + S_2}{1 - G_1 S_2}, \quad
  \mathcal{G}_2 = \frac{G_2 + S_1}{1 - G_2 S_1}.
  \label{eq:cG12-def}
\end{gather}
\end{subequations}
This expression for the SDOS incorporates the effects of diffuse scattering through $S_1$ and $S_2$.

\section{SDOS in the specular limit}
\label{sec:SDOS-specular}

Before addressing the effects of diffuse scattering, we examine the SDOS of MAES in the specular limit. In this section, we analyze the Green's functions and SDOS within the uniform gap model, in which $\Delta(y)$ is replaced everywhere by its bulk value $\Delta(\infty)$. This model is useful for elucidating the low-energy structure of the SDOS \cite{Higashitani_PRB2024}.

When $\Delta(y)$ is spatially uniform, $D(y)$ also becomes constant and is equal to $D(\infty)$. In this case, the Green's functions in the specular limit, $G_{1,2}$, can be expressed as
\begin{equation}
  G_{1,2} = -\frac{\Omega_{1,2}}{\varepsilon}, \quad
  \Omega_1 = \Omega + \Delta_{\phi}, \quad
  \Omega_2 = \Omega - \Delta_{\phi}.
\end{equation}
The poles of $G_{1,2}$ at $\varepsilon = 0$ correspond to MAES. Therefore, the Green's functions associated with the edge states (ES), excluding contributions from continuum states, can be obtained by evaluating $\Omega_{1,2}$ at $\varepsilon = 0$. This yields
\begin{equation}
  G_1^{\rm ES} = \frac{1 + s_{\phi}}{2} G^{\rm ES}, \quad
  G_2^{\rm ES} = \frac{1 - s_{\phi}}{2} G^{\rm ES},
  \label{eq:G12-ES-def}
\end{equation}
where
\begin{equation}
  G^{\rm ES} = -\frac{2|\Delta_{\phi}|}{\varepsilon}, \quad
  s_{\phi} = \frac{\Delta_{\phi}}{|\Delta_{\phi}|} = {\rm sgn}[\sin(m\phi)].
\end{equation}
The contribution from continuum states (CS) to $G_{1,2}$ is given by
\begin{equation}
  G^{\rm CS}
  = G_{1,2} - G_{1,2}^{\rm ES}
  = \frac{\varepsilon}{\Omega + |\Delta_{\phi}|}.
\end{equation}

From Eq.\ \eqref{eq:G12-ES-def}, the SDOS of MAES at a specular surface is obtained as
\begin{subequations}
  \label{eq:n12-specular}
\begin{gather}
  n_1(\phi,E) = \frac{1}{2} {\rm Im}(G_1^{\rm ES})^R
  = \frac{1 + s_{\phi}}{2}\pi |\Delta_{\phi}| \delta(E),
  \label{eq:n1-specular} \\
  n_2(\phi,E) = \frac{1}{2} {\rm Im}(G_2^{\rm ES})^R
  = \frac{1 - s_{\phi}}{2}\pi |\Delta_{\phi}| \delta(E),
  \label{eq:n2-specular}
\end{gather}
\end{subequations}
and
\begin{equation}
  n(\phi,E) = n_1(\phi,E) + n_2(\phi,E) = \pi |\Delta_{\phi}| \delta(E).
  \label{eq:n1+n2-specular}
\end{equation}

\begin{figure}
  \centering
  \includegraphics[scale=0.85]{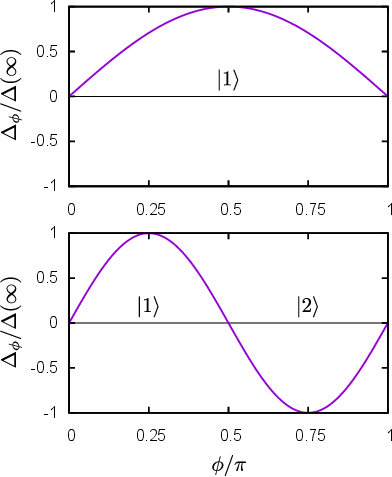}
  \caption{Angular dependence of $\Delta_{\phi}$ for (a) $p$-wave and (b) $d$-wave SC states. The $\ket{1}$ and $\ket{2}$ MAES modes exist in the regions $\Delta_{\phi} > 0$ ($s_{\phi} = +1$) and $\Delta_{\phi} < 0$ ($s_{\phi} = -1$), respectively.}
    \label{fig:Delta-ket-rel}
\end{figure}

Equations \eqref{eq:n1-specular} and \eqref{eq:n2-specular} represent the SDOS associated with the $\ket{1}$ and $\ket{2}$ MAES modes, respectively. From these expressions, or equivalently from Eq.\ \eqref{eq:G12-ES-def} (see also Fig.\ \ref{fig:Delta-ket-rel}), we identify a distinctive feature of the $p$-wave SC state: it supports only the $\ket{1}$ mode, whereas other SC states support both $\ket{1}$ and $\ket{2}$ modes. This distinction plays a key role in understanding the different responses of MAES to surface roughness in $p$- and $d$-wave SC states, as discussed in the next section.

\section{SDOS of MAES at rough surfaces}
\label{sec:SDOS-rough}

\subsection{SDOS in the $d$-wave SC state}

\begin{figure*}[t]
  \centering
  \includegraphics[scale=0.85]{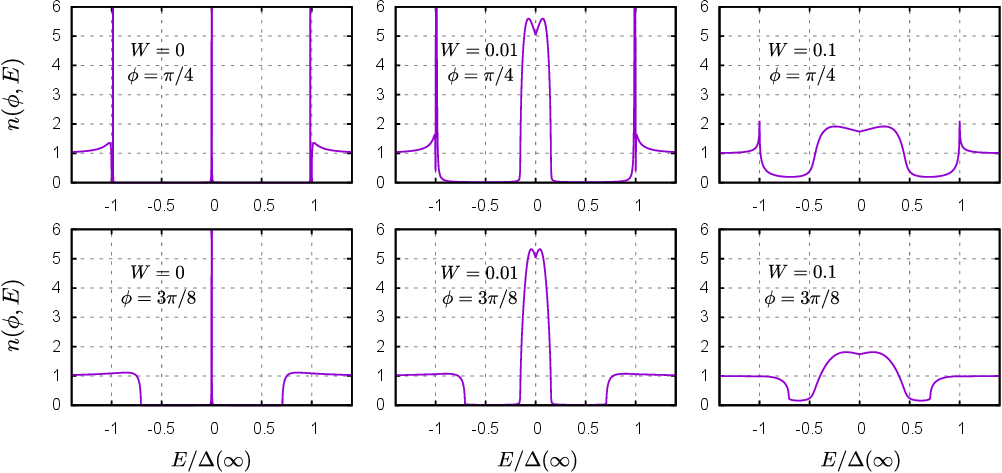}
  \caption{SDOS $n(\phi,E)$ in $d$-wave superconductors with $W = 0$ ($R = 1$), $W = 0.01$ ($R \simeq 0.92$), and $W = 0.1$ ($R \simeq 0.5$). The top panels correspond to $\phi = \pi/4$, and the bottom panels correspond to $\phi = 3\pi/8$. All results are calculated using the self-consistently determined $\Delta(y)$ shown in Fig.\ \ref{fig:sc-Delta-d-wave}.}
  \label{fig:sc-SDOS-d-wave}
\end{figure*}

\begin{figure}
  \centering
  \includegraphics[scale=0.8]{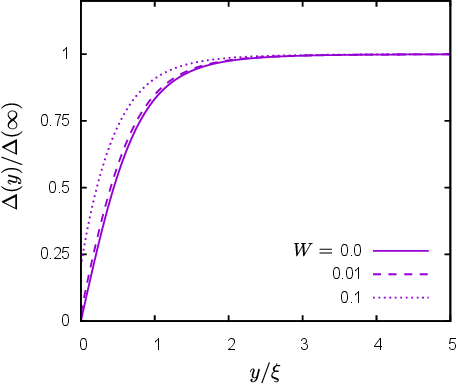}
  \caption{Spatial dependence of the self-consistent $\Delta(y)$ in $d$-wave superconductors for $W = 0$, $W = 0.01$, and $W = 0.1$. The results are calculated at $T = 0.2 T_c$, where $T_c$ is the transition temperature. The distance $y$ from the surface is scaled by the coherence length $\xi = \hbar v_F/\Delta(\infty)$.}
  \label{fig:sc-Delta-d-wave}  
\end{figure}

Figure \ref{fig:sc-SDOS-d-wave} presents numerical results for $n(\phi,E)$ in the $d$-wave SC state, obtained using the self-consistently determined $\Delta(y)$ shown in Fig.\ \ref{fig:sc-Delta-d-wave}. Due to the symmetry $\Delta_{\pi-\phi} = -\Delta_{\phi}$, the SDOS satisfies $n(\pi-\phi,E) = n(\phi,E)$. Therefore, it is sufficient to consider $\phi \in [0,\pi/2]$. As representative examples, we choose $\phi = \pi/4$ and $\phi = 3\pi/8$.

The left panels (top and bottom) show the SDOS in the specular limit for $\phi = \pi/4$ and $\phi = 3\pi/8$, respectively. At $\phi = \pi/4$, two types of edge states appear within the energy gap. One corresponds to MAES, which manifests as a zero-energy SDOS peak, as described by Eq.\ \eqref{eq:n1-specular}. The other appears near the bulk gap edge. This high-energy mode cannot be captured by the uniform gap model, as it originates from a bound state confined within the potential well formed between the surface and the suppressed $\Delta(y)$ \cite{Yamada_JPSJ1996,Sugiyama_JPSJ2020,Higashitani_PRB2024}. 

In contrast, for $\phi = 3\pi/8$, the high-energy mode is absent. This difference arises from the angular dependence of the effective potential width, which scales as $\sim \xi/\cos\theta$, where $\xi = \hbar v_F/\Delta(\infty)$ is the coherence length and $\theta = \phi - \pi/2$ is the angle measured from the surface normal. As $|\theta|$ decreases, the width becomes smaller, preventing the formation of the high-energy bound state at $\phi = 3\pi/8$.

The middle panels show the SDOS at a weakly rough surface with $W = 0.01$, corresponding to a diffuse scattering probability of approximately 8\%. Even in this weakly diffusive regime, the SDOS peak associated with MAES is significantly broadened \cite{Matsumoto_JPSJ1995,Yamada_JPSJ1996,Luck_PRB2001}. The broadening becomes more pronounced as $W$ increases, as seen in the right panels. In addition to broadening, diffuse scattering produces a characteristic V-shaped structure centered at $E = 0$ \cite{Matsumoto_JPSJ1995}. The objective of this subsection is to clarify the physical origin of this peculiar low-energy SDOS structure in the $d$-wave SC state.

In the following analysis, we examine the low-energy SDOS within the uniform gap model. For simplicity, we focus on the weakly diffusive regime ($W \ll 1$), where the MAES contribution to the SDOS is strongly localized near $E = 0$. To analyze the low-energy behavior, we neglect higher-order corrections of order $|\varepsilon/\Delta_{\phi}|^2$ in the specular-limit Green's functions $G_{1,2}$. Under this approximation, $G_1$ is expressed as
\begin{subequations}
  \label{eq:G1-low-E}
\begin{align}
  &(\phi \in [1])\quad G_1 \simeq G^{\rm ES}
  = -\frac{2|\Delta_{\phi}|}{\varepsilon},
  \label{eq:G1-low-E-a} \\
  &(\phi \in [2])\quad G_1 = G^{\rm CS}
  \simeq \frac{\varepsilon}{2|\Delta_{\phi}|},
  \label{eq:G1-low-E-b}
\end{align}
\end{subequations}
where $[1]$ and $[2]$ denote the angular regions $[0,\pi/2]$ ($s_{\phi} = +1$) and $[\pi/2,\pi]$ ($s_{\phi} = -1$), respectively. Equation \eqref{eq:G1-low-E-a} indicates that the $\ket{1}$ MAES mode dominates $G_1$ in region $[1]$, whereas Eq.\ \eqref{eq:G1-low-E-b} shows that the $\ket{1}$ MAES mode is absent in region $[2]$. Similarly, for $G_2$, we obtain
\begin{subequations}
  \label{eq:G2-low-E}
\begin{align}
  &(\phi \in [2])\quad G_2 \simeq G^{\rm ES}
  = -\frac{2|\Delta_{\phi}|}{\varepsilon}, \\
  &(\phi \in [1])\quad G_2 = G^{\rm CS}
  \simeq \frac{\varepsilon}{2|\Delta_{\phi}|}.
\end{align}
\end{subequations}

For $W \neq 0$, the SDOS is determined by the Green's functions $\mathcal{G}_{1,2}$ defined in Eq.\ \eqref{eq:cG12-def}. Using the low-energy expressions for $G_{1,2}$, we obtain
\begin{gather}
  \mathcal{G}_1
  = - \frac{1 + s_{\phi}}{2}
  \frac{2|\Delta_{\phi}| - \varepsilon S_2}
  {\varepsilon + 2|\Delta_{\phi}| S_2}
  + \frac{1 - s_{\phi}}{2}
  \frac{\varepsilon + 2|\Delta_{\phi}| S_2}
  {2|\Delta_{\phi}| - \varepsilon S_2},
  \label{eq:cG1-low-E}
  \\
  \mathcal{G}_2
  = - \frac{1 - s_{\phi}}{2}
  \frac{2|\Delta_{\phi}| - \varepsilon S_1}
  {\varepsilon + 2|\Delta_{\phi}| S_1}
  + \frac{1 + s_{\phi}}{2}
  \frac{\varepsilon + 2|\Delta_{\phi}| S_1}
  {2|\Delta_{\phi}| - \varepsilon S_1}.
  \label{eq:cG2-low-E}
\end{gather}

In the weakly diffusive limit ($W \ll 1$), we assume $|\varepsilon S_{1,2}/\Delta_{\phi}| \ll 1$ and $|S_1 S_2| \ll 1$. Under these conditions, we obtain
\begin{equation}
  \mathcal{G}_1 + \mathcal{G}_2 =
  \begin{dcases}
    \mathcal{G}_1^{\rm ES} & (\phi \in [1]), \\
    \mathcal{G}_2^{\rm ES} & (\phi \in [2]),
  \end{dcases}
\end{equation}
where
\begin{equation}
  \mathcal{G}^{\rm ES}_{1,2}
  = -\frac{2|\Delta_{\phi}|}{\varepsilon + 2|\Delta_{\phi}| S_{2,1}}.
  \label{eq:cG-ES-def}
\end{equation}
This leads to the following expression for the SDOS in the low-energy, weakly diffusive regime:
\begin{equation}
  n(\phi \in [j], E) = \frac{1}{2} {\rm Im} \left( \mathcal{G}_j^{\rm ES} \right)^R,
  \quad (j = 1,2).
  \label{eq:SDOS-d-wave-low-E-small-W}
\end{equation}

The Green's functions $\mathcal{G}^{\rm ES}_{1,2}$ play a central role in this regime. Physically, $\mathcal{G}^{\rm ES}_{1(2)}$ represents the Green's function of the $\ket{1(2)}$ MAES mode at a rough surface. The effect of diffuse scattering is incorporated through the self-energy $\Sigma_{1,2} = -2|\Delta_{\phi}| S_{2,1}$. Notably, $\Sigma_{1,2} \propto S_{2,1}$ (rather than $S_{1,2}$), indicating that the self-energy arises from inter-mode scattering between the $\ket{1}$ and $\ket{2}$ MAES modes.

The functions $S_{1,2}$ are obtained from Eq.\ \eqref{eq:S12-G}. Owing to the symmetry of the $d$-wave gap function, $\Delta_{\pi-\phi} = -\Delta_{\phi}$, we find that $S_1 = S_2$. Defining $S \equiv S_1 = S_2$, Eq.\ \eqref{eq:S12-G} reduces to
\begin{equation}
  S = W \ev{G_1^W + G_2^W}_{\phi}.
  \label{eq:S-d-wave}
\end{equation}
Applying the same low-energy and small-$W$ approximations yields
\begin{equation}
  G_1^W + G_2^W = \mathcal{G}^{\rm ES}
  = -\frac{2|\Delta_{\phi}|}{\varepsilon + 2|\Delta_{\phi}| S}.
  \label{eq:G1W+G2W-d-wave-low-E-small-W}
\end{equation}

\begin{figure}
  \centering
  \includegraphics[scale=0.8]{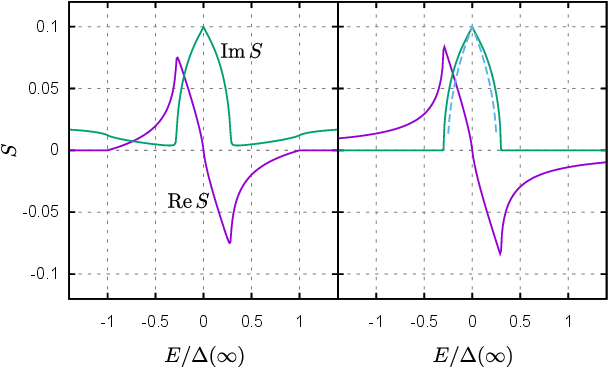}
  \caption{Energy dependence of $S$ in $d$-wave superconductors with $W = 0.01$, calculated within the uniform gap model. The left and right panels show the exact and approximate solutions of Eq.\ \eqref{eq:S-d-wave}, respectively. The approximation is obtained by substituting Eq.\ \eqref{eq:G1W+G2W-d-wave-low-E-small-W} into Eq.\ \eqref{eq:S-d-wave}. The dashed line corresponds to Eq.\ \eqref{eq:imS-d-wave-low-E-limit}.}
  \label{fig:S-d-wave}
\end{figure}

\begin{figure}
  \centering
  \includegraphics[scale=0.75]{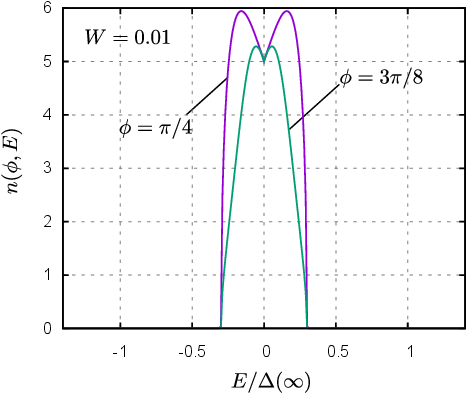}
  \caption{SDOS at $\phi = \pi/4$ and $3\pi/8$ in a $d$-wave superconductor with $W = 0.01$ ($R \simeq 0.92$). The results are obtained from Eq.\ \eqref{eq:SDOS-d-wave-low-E-small-W} using the $S$ values shown in the right panel of Fig.\ \ref{fig:S-d-wave}.}
  \label{fig:SDOS-d-wave}
\end{figure}

Figure \ref{fig:S-d-wave} shows $S$ as a function of $E/\Delta(\infty)$ for $W = 0.01$. The left panel presents the exact result obtained within the uniform gap model, while the right panel demonstrates that the low-energy behavior is well reproduced by the approximation in Eq.\ \eqref{eq:G1W+G2W-d-wave-low-E-small-W}. At $E = 0$, $S$ is purely imaginary, with $S = i\sqrt{W}$. Near $E = 0$, ${\rm Im}\,S$ exhibits a dome-shaped profile with a cusp at $E = 0$. This cusp gives rise to the V-shaped structure observed in the SDOS. Indeed, near $E = 0$, Eq.\ \eqref{eq:SDOS-d-wave-low-E-small-W} simplifies to $n(\phi,E) \simeq 1/(2\,{\rm Im}\,S)$. Figure \ref{fig:SDOS-d-wave} shows the corresponding SDOS obtained from Eq.\ \eqref{eq:SDOS-d-wave-low-E-small-W} using the $S$ values from the right panel of Fig.\ \ref{fig:S-d-wave}. The resulting low-energy structure closely reproduces the self-consistent SDOS shown in Fig.\ \ref{fig:sc-SDOS-d-wave}.

To clarify the origin of the cusp in ${\rm Im}\,S$, we rewrite Eq.\ \eqref{eq:G1W+G2W-d-wave-low-E-small-W} as
\begin{equation}
  G_1^W + G_2^W
  = -\frac{1}{S}
  \left(
    1 - \frac{\varepsilon}{\varepsilon + 2|\Delta_{\phi}| S}
  \right).
  \label{eq:G1W+GW2-rewrite}
\end{equation}
Substituting this expression into Eq.\ \eqref{eq:S-d-wave}, we obtain
\begin{equation}
  S^2 = -W (1 - \varepsilon J), \quad
  J = \frac{1}{2}\int_0^{\pi} d\phi\, \frac{\sin\phi}{\varepsilon + 2|\Delta_{\phi}| S}.
\end{equation}
The low-energy behavior of ${\rm Im}\,S$ is governed by the zero-energy limit of ${\rm Re}\,J$. To evaluate this limit, we approximate $S$ in the integrand by its zero-energy value $i\sqrt{W}$, yielding
\begin{align*}
  J
  &= \frac{1}{2}\int_0^{\pi} d\phi\,
  \frac{\sin\phi}{\varepsilon + 2|\Delta_{\phi}| i\sqrt{W}} \\
  &= -\frac{i}{4\sqrt{W}}\int_0^{\pi} d\phi\,
  \frac{\sin\phi}{|\Delta_{\phi}| - iE/(2\sqrt{W})}.
\end{align*}
From this, we obtain
\begin{align}
  {\rm Re}\,J
  &\xrightarrow{E \to 0}
  \frac{{\rm sgn}(E)\pi}{4\sqrt{W}}
  \int_0^{\pi} d\phi\, \sin\phi\, \delta(|\Delta_{\phi}|) \notag \\
  &= \frac{{\rm sgn}(E)\pi}{8\sqrt{W}\Delta(\infty)}.
  \label{eq:ReJ-ex}
\end{align}
Consequently, the low-energy behavior of ${\rm Im}\,S$ is given by
\begin{equation}
  {\rm Im}\,S = W \sqrt{1 - \frac{\pi |E|}{8\sqrt{W}\Delta(\infty)}}.
  \label{eq:imS-d-wave-low-E-limit}
\end{equation}
This expression accurately reproduces the numerical behavior of $S$, as shown by the dashed curve in Fig.\ \ref{fig:S-d-wave}. The first line of Eq.\ \eqref{eq:ReJ-ex} indicates that the cusp in ${\rm Im}\,S$ originates from inter-mode scattering near the gap node at $\phi = \pi/2$ within the broadened MAES flat band.

\subsection{SDOS in the $p$-wave SC state \label{sec:sdos-p-wave}}

The low-energy expressions in Eqs.\ \eqref{eq:cG1-low-E} and \eqref{eq:cG2-low-E} for $\mathcal{G}_{1,2}$ can also be applied to the $p$-wave SC state. The key difference from the $d$-wave case arises from the behavior of $s_{\phi}$. In the $p$-wave case, $s_{\phi} = +1$ over the entire range of $\phi$. Consequently, $\mathcal{G}_{1,2}$ reduce to
\begin{equation}
  \mathcal{G}_1
  = -\frac{2|\Delta_{\phi}| - \varepsilon S_2}
  {\varepsilon + 2|\Delta_{\phi}| S_2}, \quad
  \mathcal{G}_2 =
  \frac{\varepsilon + 2|\Delta_{\phi}| S_1}
  {2|\Delta_{\phi}| - \varepsilon S_1}.
  \label{eq:cG1-cG2-p-wave}
\end{equation}

The Green's functions $G_{1,2}^{W}$ correspond to $\mathcal{G}_{1,2}$ without the second term in the numerator. Accordingly, $S_{1,2}$ are determined by
\begin{subequations}
  \label{eq:S12-p-wave}
\begin{gather}
  S_1 = 2W \ev{
      -\frac{2|\Delta_{\phi}|}{\varepsilon + 2|\Delta_{\phi}| S_2}
    }_{\phi},
  \label{eq:S1-p-wave} \\
  S_2 = 2W \ev{
      \frac{\varepsilon}{2|\Delta_{\phi}| - \varepsilon S_1}
    }_{\phi}.
  \label{eq:S2-p-wave}
\end{gather}
\end{subequations}
These three equations form a closed set that determines the SDOS of MAES in the $p$-wave SC state for arbitrary $W$.

\begin{figure}
  \centering
  \includegraphics[scale=0.8]{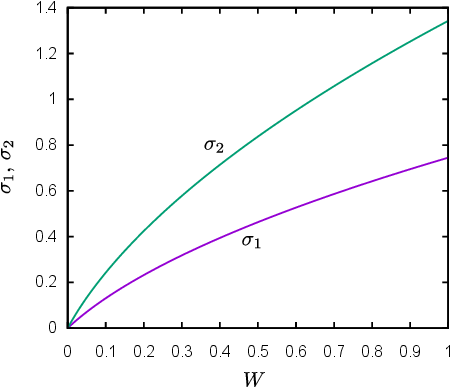}
  \caption{Dependence of $\sigma_{1,2}$ on $W$ in the $p$-wave SC state, obtained from Eq.\ \eqref{eq:sigma12-W}.}
  \label{fig:sigma12-p-wave}
\end{figure}

The solution of Eq.\ \eqref{eq:S12-p-wave} exhibits the following energy dependence:
\begin{equation}
  S_1 = -\frac{2\Delta(\infty)}{\varepsilon} \sigma_1, \quad
  S_2 = \frac{\varepsilon}{2\Delta(\infty)} \sigma_2.
  \label{eq:S12-sigma12}
\end{equation}
Here, $\sigma_{1,2}$ are energy-independent parameters that satisfy
\begin{subequations}
  \label{eq:sigma12-W}
\begin{gather}
  \sigma_1 = 2W \ev{\frac{\sin\phi}{1 + \sigma_2 \sin\phi}}_{\phi},
  \label{eq:sigma1-W} \\
  \sigma_2 = 2W \ev{\frac{1}{\sin\phi + \sigma_1}}_{\phi}.
  \label{eq:sigma2-W}
\end{gather}
\end{subequations}
These parameters determine the dependence of the SDOS on $W$. Numerical evaluation of Eq.\ \eqref{eq:sigma12-W} shows that both $\sigma_1$ and $\sigma_2$ increase monotonically with $W$ (Fig.\ \ref{fig:sigma12-p-wave}). Substituting Eq.\ \eqref{eq:S12-sigma12} into Eq.\ \eqref{eq:cG1-cG2-p-wave} yields
\begin{align}
  \mathcal{G}_1 &=
  \frac{1}{1 + \sigma_2 \sin\phi}
  \left(
    -\frac{2|\Delta_{\phi}|}{\varepsilon}
    + \frac{\varepsilon \sigma_2}{2\Delta(\infty)}
  \right), \\
  \mathcal{G}_2 &=
  \frac{1}{\sin\phi + \sigma_1}
  \left(
    \frac{\varepsilon}{2\Delta(\infty)}
    - \frac{2|\Delta_{\phi}| \sigma_1}{\varepsilon}
  \right).
\end{align}
These expressions lead to the SDOS
\begin{equation}
  n(\phi,E) = Z(\phi)\, \pi |\Delta_{\phi}| \delta(E),
  \label{eq:SDOS-p-wave}
\end{equation}
where
\begin{equation}
  Z(\phi)
  = \frac{1}{1 + \sigma_2 \sin\phi}
  + \frac{\sigma_1}{\sin\phi + \sigma_1}.
\end{equation}

Equation \eqref{eq:SDOS-p-wave} demonstrates the robustness of MAES in the $p$-wave SC state. Even at rough surfaces, MAES retain a sharp zero-energy peak in the SDOS, and the effect of diffuse scattering is fully captured by the renormalization factor $Z(\phi)$. The total spectral weight, $\int d\phi\, Z(\phi)\pi |\Delta_{\phi}|$, remains unchanged by diffuse scattering. Indeed, the deviation of the total spectral weight from the specular case is given by
\begin{align*}
  &\int_0^{\pi} d\phi\, Z(\phi)\pi |\Delta_{\phi}|
  - \int_0^{\pi} d\phi\, \pi |\Delta_{\phi}| \\
  &= \int_0^{\pi} d\phi\,
  \left(
    \frac{1}{1 + \sigma_2 \sin\phi} - 1
    + \frac{\sigma_1}{\sin\phi + \sigma_1}
  \right)\pi |\Delta_{\phi}| \\
  &\propto
  -\sigma_2 \ev{\frac{\sin\phi}{1 + \sigma_2 \sin\phi}}_{\phi}
  + \sigma_1 \ev{\frac{1}{\sin\phi + \sigma_1}}_{\phi}.
\end{align*}
Equation \eqref{eq:sigma12-W} ensures that the final expression vanishes.

The robustness of MAES in the $p$-wave state can also be understood qualitatively in the weakly diffusive regime. For $W \ll 1$, Eq.\ \eqref{eq:S12-p-wave} simplifies to
\begin{subequations}
\begin{gather}
  S_1
  = 2W \ev{
      -\frac{2|\Delta_{\phi}|}{\varepsilon}
    }_{\phi}
  = 2W \ev{G^{\rm ES}}_{\phi}, \\
  S_2
  = 2W \ev{
      \frac{\varepsilon}{2|\Delta_{\phi}|}
    }_{\phi}
  = 2W \ev{G^{\rm CS}}_{\phi}.
\end{gather}
\end{subequations}
Thus, $S_1$ and $S_2$ describe MAES–MAES and MAES–CS scattering processes, respectively. In the $p$-wave SC state, where the $\ket{2}$ MAES mode is absent, MAES–MAES scattering corresponds to intra-mode scattering within the flat band of the $\ket{1}$ MAES mode. In contrast to the inter-mode scattering between $\ket{1}$ and $\ket{2}$ modes in the $d$-wave state, this intra-mode scattering does not lead to broadening of the MAES SDOS. Furthermore, MAES–CS scattering results in $S_2 \propto \varepsilon$, implying that the self-energy $\Sigma_1 = -2|\Delta_{\phi}| S_2$ of the $\ket{1}$ mode has no imaginary component and therefore does not contribute to SDOS broadening.

\section{Conclusions}

We have investigated the effects of surface roughness on the flat band of MAES in $p$- and $d$-wave superconducting (SC) states. As established in previous theoretical studies, the response of MAES to surface roughness differs markedly between these two SC states. In the $d$-wave state, diffuse surface scattering significantly broadens the SDOS of MAES \cite{Matsumoto_JPSJ1995,Yamada_JPSJ1996,Luck_PRB2001}. Moreover, this broadening is accompanied by an unusual V-shaped structure centered at zero energy \cite{Matsumoto_JPSJ1995}. In contrast, the SDOS of MAES in the $p$-wave state remains highly robust against diffuse surface scattering \cite{Nagato_JLTP1998,Bakurskiy_PRB2014}. To clarify the physical origin of this pronounced difference, we analyzed the SDOS using the random S-matrix theory \cite{Higashitani_PRB2024,Nagato_JLTP1996,Yamada_JPSJ1996}.

We demonstrated that the fragility of MAES in the $d$-wave state arises from the presence of two distinct MAES modes occupying different regions of the Fermi surface. Similar edge modes appear in two-dimensional (2D) chiral superconductors, where they are explicitly visible in the dispersion relation as a function of momentum parallel to the surface \cite{Huang_PRB2014,Tada_PRL2015,Wang_PRB2018,Sugiyama_JPSJ2020}. In contrast, the modes in the flat band of MAES are not directly resolved in the dispersion and must instead be identified through analysis of the wave function or Green's function. In our previous study of 2D chiral superconductors \cite{Higashitani_PRB2024}, we showed that mode mixing induced by diffuse scattering leads to substantial broadening of the edge-state SDOS. The present results demonstrate that a similar mechanism operates for the flat band. However, a qualitative difference emerges between chiral and non-chiral superconductors. In chiral systems, SDOS broadening exhibits a conventional Lorentzian profile \cite{Higashitani_PRB2024}, whereas in non-chiral systems the broadened peak develops into a V-shaped profile. We attribute this distinctive feature to the presence of gap nodes: a full gap exists over the Fermi surface in 2D chiral superconductors, whereas nodes are present in non-chiral superconductors. Diffuse scattering near these gap nodes gives rise to the V-shaped structure.

In contrast to the $d$-wave case, the edge state in the $p$-wave state consists of a single MAES mode, and mode mixing is therefore intrinsically absent. This absence underlies the robustness of MAES in the $p$-wave state. We have shown that MAES manifests in the SDOS as a sharp zero-energy peak even in the presence of diffuse scattering, with the effect of scattering incorporated solely through a renormalization factor of the zero-energy spectral weight. These findings offer an alternative perspective on the robustness of MAES, distinct from explanations based on odd-frequency pairing \cite{Higashitani_PRB2012,Bakurskiy_PRB2014,Miyawaki_PRB2018}.

\begin{acknowledgments}
  This work was partly supported by JSPS KAKENHI through Grant No. 22K03530.
\end{acknowledgments}

\appendix*

\section{Derivation of Eq.\ \eqref{eq:SDOS-formula}}
\label{sec:appendix}

Combining Eqs.\ \eqref{eq:gp-DF}, \eqref{eq:gm-DF}, and
\eqref{eq:gs-braket}, we obtain
\begin{equation}
  \hat{\rho}_{3}[\hat{g}_\alpha(0) - \hat{g}_S^{}]
  = A [G_1 \hat{u}_{11} - G_2 \hat{u}_{22} + \alpha i(\hat{u}_{21}-\hat{u}_{12})],
  \label{eq:appen:g-gs}
\end{equation}
where
\begin{equation}
  A = i\frac{F(0)-D(0)}{1+D(0)F(0)},\ \ \hat{u}_{nn'} = \ketbra{n}{n'}.
\end{equation}
Using Eq.\ \eqref{eq:bcon-F} to derive the surface value $F(0)$
\begin{align}
  F(0)
  &= \frac{\mathbb{S}_1\mathbb{S}_2D(0)
    + i(\mathbb{S}_1-\mathbb{S}_2)/2}{1 + i(\mathbb{S}_1-\mathbb{S}_2)D(0)/2}
    \notag \\
  &= \frac{i\mathbb{S}_1[1-i\mathbb{S}_2D(0)]
    - i\mathbb{S}_2(1+i\mathbb{S}_1D(0))}
    {[1-i\mathbb{S}_2D(0)] + [1+i\mathbb{S}_1D(0)]},
\end{align}
where
\begin{equation}
  \mathbb{S}_{1,2} = \frac{1 + iS_{1,2}}{1 - iS_{1,2}}.
\end{equation}
Thus, $A$ can be written as
\begin{align}
  A &= \frac{1}{2}
      \left(
        \frac{\mathbb{S}_2-iD(0)}{1-i\mathbb{S}_2D(0)} -
        \frac{\mathbb{S}_1+iD(0)}{1+i\mathbb{S}_1D(0)}
      \right) \notag\\
    &= \frac{1}{1-G_1S_2} - \frac{1}{1-G_2S_1}.
      \label{eq:appen:A-ex}
\end{align}
Substituting Eq. \eqref{eq:appen:A-ex} into Eq.\ \eqref{eq:appen:g-gs} yields
\begin{align*}
  \hat{\rho}_{3}\hat{g}_\alpha(0)
  &= \left( \frac{1}{1-G_1S_2} - \frac{G_2S_1}{1-G_2S_1} \right)G_1\hat{u}_{11}
  \notag \\
  &+ \left( \frac{1}{1-G_2S_1} - \frac{G_1S_2}{1-G_1S_2} \right)G_2\hat{u}_{22}
  \notag \\
  &+ \alpha i\left( \frac{1}{1-G_1S_2} - \frac{1}{1-G_2S_1} \right)
    (\hat{u}_{21} - \hat{u}_{12}).
\end{align*}
Using $G_1G_2 = -1$, we obtain
\begin{align*}
  \hat{\rho}_{3}\hat{g}_\alpha(0)
  &= \left( \frac{G_1}{1-G_1S_2} + \frac{S_1}{1-G_2S_1} \right)\hat{u}_{11}
  \notag \\
  &+ \left( \frac{G_2}{1-G_2S_1} + \frac{S_2}{1-G_1S_2} \right)\hat{u}_{22}
  \notag \\
  &+ \alpha i\left( \frac{1}{1-G_1S_2} - \frac{1}{1-G_2S_1} \right)
    (\hat{u}_{21} - \hat{u}_{12}).
\end{align*}
Finally, the SDOS formula \eqref{eq:SDOS-formula} is derived by substituting ${\rm Tr}[\hat{\rho}_{3}\hat{g}_\alpha(0)]$ into Eq.\ \eqref{eq:SDOS-def}.

%

\end{document}